# TRACTOGRAPHY-BASED PARCELLATION OF CEREBELLAR DENTATE NUCLEI VIA A DEEP NONNEGATIVE MATRIX FACTORIZATION CLUSTERING METHOD

*Xiao Xu[1, 2], Yuqian Chen[3, 4], Leo Zekelman[3], Yogesh Rathi[3], Nikos Makris[3], Fan Zhang[3], and Lauren J. O'Donnell[3]*

[1] University of Electronic Science and Technology of China, Chengdu, China
[2] University of Glasgow, Glasgow, UK
[3] Harvard Medical School, MA, USA
[4] The University of Sydney, NSW, Australia

## ABSTRACT

As the largest human cerebellar nucleus, the dentate nucleus (DN) functions significantly in the communication between the cerebellum and the rest of the brain. Structural connectivity-based parcellation has the potential to reveal the topography of the DN and enable the study of its subregions. In this paper, we investigate a deep nonnegative matrix factorization clustering method (DNMFC) for parcellation of the human DN based on its structural connectivity using diffusion MRI tractography. We propose to describe the connectivity of the DN using a set of curated tractography fiber clusters within the cerebellum. Experiments are conducted on the diffusion MRI data of 50 healthy adults from the Human Connectome Project. In comparison with state-of-the-art clustering methods, DN parcellations resulting from DNMFC show better quality and consistency of parcels across subjects.

## 1. INTRODUCTION

The dentate nucleus (DN) is critical for communication between the cerebellum and the rest of the brain. Located in each cerebellar hemisphere and farthest from the cerebellar midline, the DN is the largest cerebellar nucleus [1]. The DN participates in both motor and non-motor functions, contributing to sensorimotor processes and higher cerebellar functions [2], [3].

Few previous studies have focused on the parcellation of the DNs using magnetic resonance imaging (MRI). One study utilized diffusion MRI (dMRI) probabilistic tractography to perform parcellation of the human DN, resulting in two parcellated zones: motor rostrodorsal and non-motor ventro-caudal, suggesting the existence of different functional units within the DN [3]. Another study used a whole-brain MRI based multimodal approach to obtain three DN atlases, two of which were based on constrained spherical deconvolution tractography with the third atlas derived from a fuzzy C-means clustering method considering dMRI microstructural properties [4]. This study also demonstrated two coherently distinct regions responsible for motor and non-motor functions. In addition to these MRI studies, a cadaver dissection approach demonstrated four regions of the dentate nucleus according to their connections with the superior cerebellar peduncle [5]. Overall, these studies highlight the potential for DN parcellation of using information about its connectivity.

Here we propose a new strategy for DN parcellation based on dMRI tractography that has been parcellated into fiber clusters. This strategy performs brain parcellation based on the fiber clusters that intersect each voxel [6], [7]. In this work, we leverage an atlas of fiber clusters that has been successfully used for consistent white matter parcellation across the lifespan [8]. Previous work has demonstrated the potential of brain parcellation by clustering information about fiber clusters that intersect each voxel [6]. Here we investigate the potential of such a strategy to finely parcellate deep nuclei such as the dentate. We propose to cluster dentate voxels based on their connectivity, as described by the tractography clusters that pass through each voxel.

In terms of methodology, given the fact that there is yet no ground truth for a fine-scale parcellation of the DN, unsupervised learning is a promising direction to cluster the voxels within the DN for parcellation. Deep Convolutional Embedded Clustering (DCEC) [9] is a widely used method for unsupervised learning tasks. This method combines convolutional autoencoders (CAE) for embedding and k-means for clustering, within a deep learning framework. As an alternative to k-means, which is highly sensitive to initial conditions, sparse nonnegative matrix factorization (NMF) has been shown to give more consistent clustering solutions [10]. In this work, we investigate improvements to DCEC by replacing the k-means algorithm in the embedded clustering layer with NMF. Our methodological investigation is motivated by related work demonstrating non-deep NMF clustering for brain parcellation [6] and several recent studies implementing nonlinear NMF via combination with autoencoders [11], [12].

In this work, our main contribution is to investigate a deep NMF clustering method for parcellation of the human DN based on a novel description of its structural connectivity. We assess DN parcellation results across multiple subjects and in comparison with the DCEC and NMF clustering methods.

## 2. MATERIALS AND METHODS

### 2.1. Dataset and preprocessing

dMRI data of 50 healthy adults from the Human Connectome Project (HCP) (a subset of the "100 Unrelated subjects" release) [13] was utilized. This data was processed with the HCP minimal processing pipeline, including motion and distortion artifact correction and coregistration to the standard MNI space [14]. We used 40 subjects' data for training and 10 for testing. Subject-specific DN masks were obtained by applying the spatially unbiased atlas template of the cerebellum (SUIT) method [15]. Subject-specific unscented Kalman filter tractography (UKF) tractography [16] was performed, followed by tractography parcellation using an anatomically curated white matter atlas that includes 800 fiber clusters [17].

We first annotated each voxel in the DN with information about intersecting fiber clusters, and then we used this annotation as feature input to our deep network. For each voxel in the DN mask, indices of intersecting fiber clusters were recorded. We focused on a set of curated, bilaterally defined [7], clusters that intersected cerebellar cortex and the DNs. In general, these fiber clusters enter the DNs from the cerebellar cortex and exit the cerebellum via the superior cerebellar peduncle (**Fig. 1**). dMRI tractography visualization was performed in 3D Slicer (www.slicer.org) via SlicerDMRI (dmri.slicer.org) [18], [19].

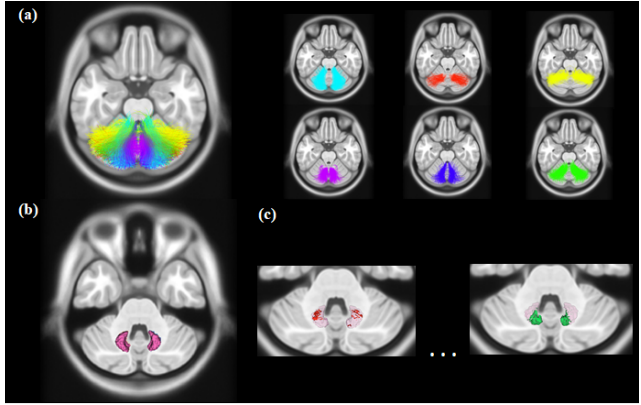

**Fig. 1.** (a) Six anatomically curated fiber clusters intersecting the DNs. (b) Mask of the DNs. (c) Regions of the DNs intersected by two example fiber clusters.

### 2.2. Voxel annotation

We first compute a feature representation for each voxel within the DN mask. Specifically, for each voxel $i$, a feature vector $v_i = (x_j | j = 1, …, 6)$ is computed, where $x_j$ is a binary value indicating whether the voxel is intersected with fiber cluster $j$. In our study, we consider a voxel is intersected with a cluster if at least one fiber in the cluster passes through the voxel.

### 2.3. Deep NMF clustering (DNMFC)

As is shown in **Fig. 2**, the proposed DNMFC is based on the network structure of DCEC, consisting of a CAE to learn an embedding feature for each input voxel and a clustering layer to assign a cluster label for each voxel. However, the K-means clustering method in the embedded layer is replaced with NMF. Here, CAE is updated by minimizing the total loss ($L$) that combines the reconstruction loss ($L_r$) and the clustering loss ($L_c$):

$$L = L_r + \gamma L_c \quad (1)$$

where γ is a coefficient controlling the degree of distorting the embedded space.

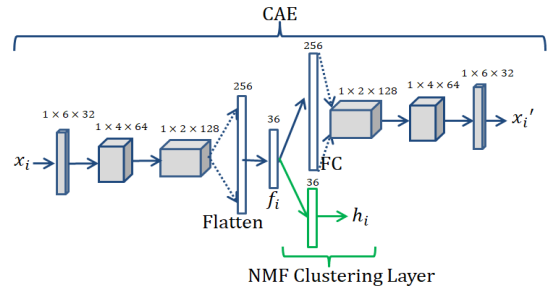

**Fig. 2.** Network structure of DNMFC.

#### 2.3.1. CAE and reconstruction loss

CAE is preserved from DCEC and aims to extract an embedding feature from the input feature vector $x_i$ (as shown in **Fig. 2**). However, we improve network architecture to learn more latent features useful to the embedded NMF clustering layer. To do so, we use a fully connected layer that maps the flattened feature vector to a higher dimensional embedding vector $f_i$ ($D$=36). This is different from the original DCEC, where $f_i$ usually has the same dimension as $x_i$ ($D$=6). Then, the rest of the network and the reconstruction loss are the same as the DCEC [9].

#### 2.3.2. NMF clustering layer and clustering loss

Here for the input of N total voxels, the transpose of the learned feature matrix $F \in R^{36 \times N}$ is decomposed by NMF into a component matrix $W \in R^{36 \times K}$ and a coefficient matrix $H \in R^{K \times N}$ ($F \approx WH$), where $K$ is the total number of parcels. Here each column of $W$ is seen as a set of features for a centroid and $W^{-1}F$ can give an approximate value of $H$. Each column of $H$ direcly gives $K$ scores (soft labels) for $K$ parcels and the maximum score determines which parcel the voxel belongs to. The NMF problem can be described mathematically as:

$$\min_{W \geq 0, H \geq 0} \frac{1}{2}||F - WH||_F^2 \quad (2)$$

where $||\cdot||_F$ is the Frobenius norm.

Unlike K-means in DCEC, where soft labels are obtained by mapping through a student's t-distribution, $H$ in NMF directly maps the learned feature vector $f_i$ from CAE to a soft label $h_i$ in the embedded clustering layer. Thus $h_{ik}$ is the $k_{th}$ entry of $h_i$, representing the probability of $f_i$ belonging to the $k_{th}$ parcel. Then the soft labels $h_i$ are directly used in the DCEC clustering loss:

$$L_c = KL(P||H) = \sum_i \sum_k p_{ik} log \frac{p_{ik}}{h_{ik}} \quad (3)$$

where $KL(\cdot)$ is the Kullback-Leibler divergence and $P$ is the target distribution defined in [9].

*2.3.3. Implementation details*
We implement our method using Python and Keras. As is shown in **Fig. 2**, three convolution layers use 32, 64 and 128 filters, respectively. Kernel sizes are $1 \times 6$, $1 \times 4$ and $1 \times 2$ with stride length equal to 2. Identical sizes are used for deconvolutional layers. All the layers are activated by the ReLu function to preserve the non-negativity of features. CAE is pretrained for 200 epochs with $\gamma$ set as 0, and the component matrix $W$ and target distribution $P$ are initialized based on embedded features. Then $\gamma$ is set as 0.1 to update the CAE parameters and $W$. Once the change of label assignments between two consecutive updates is less than a threshold $\delta = 0.1\%$, the training process ends. In a final step, to correct isolated voxels, a median filter is applied to the clustering results.

## 3. RESULTS AND DISCUSSION

### 3.1. Evaluation metrics

The quality of the parcellation result was evaluated with two metrics. We used the Silhouette coefficient ($S$) to evaluate the cohesion within parcels and the separation between parcels, where a higher S indicates a better quality of parcels [20]. Computation was done based on Scikit-learn [21]. The other metric was the Sørensen–Dice coefficient ($Dice$) used for evaluating the spatial consistency of parcellation results across subjects [22]. For a pair of subjects, their overlapping degree of the $k_{th}$ parcel is defined as $Dice_{k\text{-}pair}$:

$$Dice_{k-pair} = \left(\frac{2|L_i \cap L_j|}{|L_i|+|L_j|}\right)_{i \neq j}, \quad k = 1, ..., K \quad (4)$$

where $L_i$ and $L_j$ are the label maps for the $i_{th}$ and $j_{th}$ subject. The mean value of $Dice_{k\text{-}pair}$ from all the pairs was then obtained as $Dice_k$. A higher Dice score indicates stronger consistency of parcellation across subjects. We determined the number of DN clusters, $K$, by testing a range of values larger than 2, where $K=3$ gave the best Dice performance.

### 3.2. Comparison across methods

The results in **Table 1** indicate that the proposed DNMFC method produces the best quality dentate parcellation results. The silhouette coefficient results indicate that the within-cluster cohesion and between-cluster separation of the DNMFC method are higher than those of the compared DCEC and NMF methods. The Dice overlap metrics of Parcel 1 are similar across methods (as this parcel is initialized to contain voxels intersected by only one or fewer streamline points). The Dice overlap metrics of Parcels 2 and 3 demonstrate the high performance of DNMFC.

**Table 1.** Comparison of 3 parcellation methods

|       | $S$ | $Dice_1$ | $Dice_2$ | $Dice_3$ | $Dice_{mean}$ |
|-------|--------|--------|--------|--------|--------|
| DNMFC | **0.3089** | 0.6499 | **0.3053** | **0.4160** | **0.4571** |
| DCEC  | 0.2798 | 0.6519 | 0.0229 | 0.3967 | 0.3571 |
| NMF   | 0.1002 | **0.6520** | 0.1394 | 0.1818 | 0.3244 |

### 3.3. Dentate Parcellation

Visualization of the DN parcellation is provided in **Fig. 3** and **Fig. 4**. We observe three parcels. Parcel 1 is medial and inferior, and generally is less intersected by the fiber clusters of the current atlas. Parcel 2 is superior and lateral, while Parcel 3 is superior and medial/posterior.

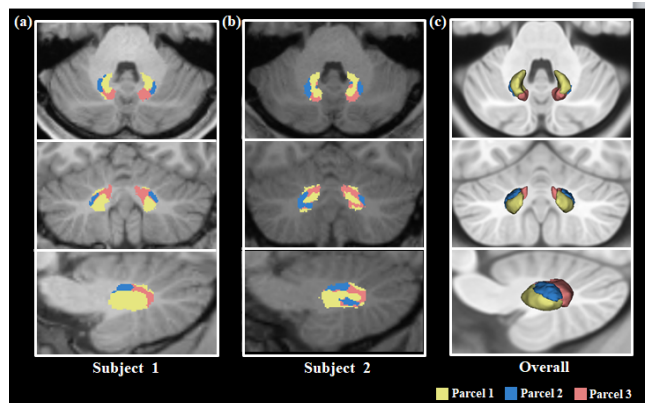

**Fig. 3.** Visualization of parcellation results in two example subjects (a,b) and overall result in 10 testing subjects (c).

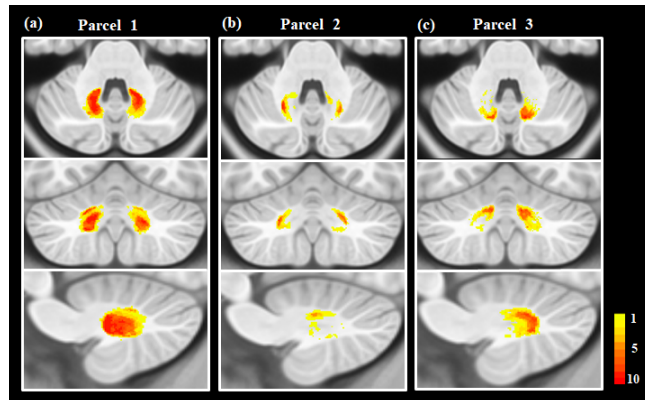

**Fig. 4.** Heatmaps for three parcels across the 10 testing subjects. Darker red color indicates higher overlap across subjects. Heatmaps for Parcel 1 (a), 2 (b), and 3 (c).

Previous work in dentate parcellation focused on motor vs non-motor regions, and demonstrated motor-related connectivity of the superior dentate nucleus [3], [4]. However, the fiber clusters in the employed tractography atlas (**Fig. 1**) connect mainly to the posterior (non-motor) cerebellum. (Shorter streamlines connecting DNs and anterior cerebellar cortex are not included in the employed tractography fiber cluster atlas.) Thus our current results subdivide regions within the dentate that have different connectivities to the posterior cerebellum, which has primarily cognitive function. Future work may incorporate a more detailed, cerebellum-specific tractography atlas to investigate the potential for finer dentate parcellation.

## 4. CONCLUSION

In this paper, we utilized intersecting fiber cluster information to parcellate an important structure in the human cerebellum, the dentate nucleus. Six fiber clusters with strong anatomical meaning were selected for voxel annotation. To perform dentate parcellation, we applied an improved DCEC method with the NMF algorithm incorporated in the embedded clustering layer. Experimental results illustrate good quality of parcels and consistency of parcellation across subjects. The ability to perform fine tractography-based parcellation may provide insights into the detailed structure and function of the dentate.

## 5. COMPLIANCE WITH ETHICAL STANDARDS

This study was conducted retrospectively using public HCP imaging data [13]. No ethical approval was required.

## 6. ACKNOWLEDGEMENTS

We acknowledge the following NIH grants: P41EB015902, R01MH125860, R01MH119222 and R01NS125781.